\documentclass{epl}
\usepackage{amsmath}
\usepackage{graphicx}

\graphicspath{{figures/}}

\usepackage{units,xspace}
\newcommand{\micron}{\ensuremath{\unit{\mu m}}\xspace}
\renewcommand{\vec}[1]{\ensuremath{\mathbf{#1}}}
\newcommand{\vecr}{\vec{r}}
\newcommand{\gw}{\gamma_0^w\left(\frac{a}{h}\right)}
\newcommand{\order}[1]{{\mathcal O}\left(#1\right)}

\begin{document}

\title{Colloidal hydrodynamic coupling in concentric optical vortices}
\shorttitle{Drag in concentric optical vortices}

\author{Kosta Ladavac\inst{1} and David G. Grier\inst{2}}
\institute{
  \inst{1} James Franck Institute and Dept. of Physics,
The University of Chicago, Chicago, IL 60637\\
  \inst{2} Dept. of Physics and Center for Soft Matter Research,
New York University, New York, NY 10003
}
\shortauthor{Ladavac and Grier}

\date{\today}

\maketitle

\begin{abstract}
  Optical vortex traps created from helical modes of light can drive
  fluid-borne colloidal particles in circular trajectories.
  Concentric circulating rings of particles formed by coaxial optical
  vortices form a microscopic Couette cell, in which the amount of
  hydrodynamic drag experienced by the spheres depends on the relative
  sense of the rings' circulation.
  Tracking the particles' motions makes possible measurements
  of the hydrodynamic coupling between the circular particle trains and
  addresses recently proposed hydrodynamic instabilities
  for collective colloidal motions on optical vortices.
\end{abstract}

A beam of light with helical wavefronts \cite{allen92} focuses to a ring-like optical trap
known as an optical vortex \cite{he95,simpson96,gahagan96}, 
which not only traps micrometer-scale objects,
but also exerts torque on them \cite{he95a,friese96,simpson97}.
Since their introduction nearly a decade ago, optical vortices have
been used to probe the nature of photon orbital angular momentum 
\cite{friese96,simpson97,oneil00,oneil02,curtis03}
and to create microoptomechanical devices such as rotary pumps and
mixers 
\cite{curtis02,ladavac04a}.
In this paper, we demonstrate a new class of microoptomechanical devices
resembling Couette cell rheometers that are based on optimally matched pairs of 
concentric optical vortices \cite{guo04} created with the holographic optical
tweezer technique \cite{dufresne98,reicherter99,liesener00,dufresne01a,curtis02}.
These paired-vortex machines constitute a useful model system for studying
many-body colloidal hydrodynamics.

The field, $\psi(\vecr) = u(r) \, \exp( i \ell \theta )$,
in an optical vortex is characterized by a transverse
phase profile $\varphi_\ell(\vecr) = \ell \theta$, where 
$\vecr = (r,\theta)$ is a polar coordinate in a plane normal to the optical
axis and $\ell$ is an integer winding number, also known as a
topological charge, that characterizes the beam's helicity.
The amplitude profile $u(r)$ typically is that of a conventional TEM$_{00}$ laser
beam whose wavefronts are imprinted with $\varphi_\ell(\vecr)$ to create the helical mode.
In our implementation, a liquid crystal spatial light modulator
(Hamamatsu X7550 PAL-SLM) encodes computer-generated holograms on the
wavefronts of a $\lambda = 532~\unit{nm}$ 
beam from a frequency-doubled diode-pumped Nd:YVO$_4$ laser
(Coherent Verdi).
The modulated beam is relayed to a microscope
objective lens (Zeiss 100$\times$ NA 1.4 oil immersion S-Plan Apochromat) mounted in
an inverted optical microscope (Zeiss S100-TV), which focuses
it into an optical vortex.

The same objective lens can be used to create images of objects interacting with
the projected optical vortex.  Provided that care is taken to minimize or correct
for aberrations in the optical train, this system is capable of trapping one or more
colloidal particles and circulating them at up to several hundred revolutions per minute.

\begin{figure}[t!]
  \centering
  \includegraphics[width=0.75\columnwidth]{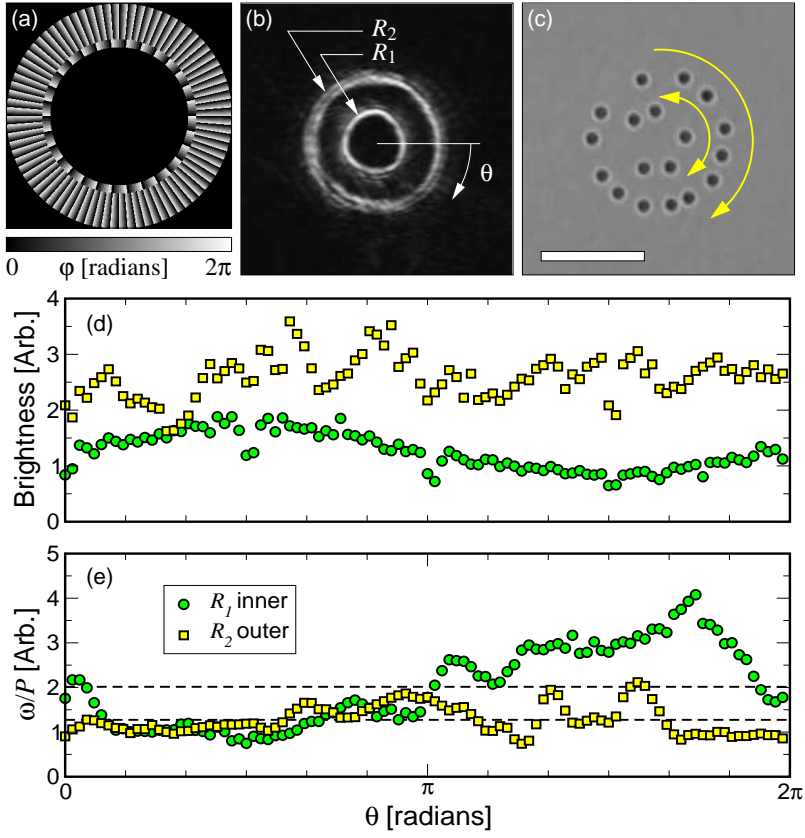}
  \caption{Creating concentric vortices from a beam of light.
    (a) Gray-scale representation of the phase hologram, 
    $\varphi(\vec{r})$, encoding two concetric vortices of 
    topologicl charges $l_1 = -30$ and $l_2 = 80$.
    (b) Focused image of the vortices projected by $\varphi(\vec{r})$,
    after aberration correction.
    (c) Bright-field image of $2a=0.99~\micron$
    diameter polystyrene spheres trapped and rotating
    in optimized optical vortices.
    The scale bar indicates $10\micron$.
    (d) Measured brightness of the laser light around the
    circumference of the rings in (b).
    (e) Circumferential speeds of an individual colloidal sphere,
    measured separately on the inner and outer rings, and normalized
    by laser power.
 }
  \label{fig:setup}
\end{figure}

If the radial amplitude profile $u(r)$ varies slowly across the 
optical train's aperture, $A$,
then the focused optical vortex takes the form of a ring of radius
\cite{curtis03,sundbeck04}
\begin{equation}
  R_\ell \approx \frac{\lambda f}{A} \, \left( 1 + \frac{\ell}{\ell_0} \right),
  \label{eq:radius}
\end{equation}
surrounded by a hierarchy of concentric diffraction rings.
Here, $f$ is the objective lens' focal length and $\ell_0 \approx 10$ arises
from the SLM's pixellated structure \cite{curtis03}.

Very recently, Guo \emph{et al.} \cite{guo04} showed
that the extraneous diffraction rings can
be ascribed to rays of light emanating from the central region of the
phase mask, 
$\varphi_\ell(\vec{r})$, while
the principal ring is projected from the outer region.
This explains why optical vortices with
winding numbers exceeding $\ell = 200$ can be projected with a
pixellated SLM \cite{curtis02,curtis03,sundbeck04}
even though features in the associated phase mask exceed the device's
Nyquist wavenumber near the optical axis.
It also suggests that the generally undesirable 
diffraction rings can be suppressed
by imposing an annular radial profile $u(r) = \Theta(r - r_1)
\Theta(r_0 - r)$ on the source beam,
where $\Theta(x)$ is the Heaviside step function.
The outer limit, $r_1$, replaces 
the system's aperture $A$ in Eq.~(\ref{eq:radius}) to set 
the optical vortex's radius, and the inner limit, $r_0$, can be adjusted to
optimize the optical vortex's radial profile.
Following Ref.~\cite{guo04}, the optimal inner limit
that eliminates diffraction rings without degrading the vortex's
principal ring is approximately given by
\begin{equation}
  \label{eq:optimal}
  r_0^c \approx \frac{2j'_{l,1}}{j'_{l,2}+j'_{l,3}}\, r_1,
\end{equation}
where $j^\prime_{\ell,n}$ is the $n$-th zero of the derivative of the Bessel function
$J_\ell(x)$.
Guo \emph{et al.} \cite{guo04} point out that the principal ring's intensity
is diminished when $r_0$ exceeds $r_0^c$, although its radius remains unchanged.

The unused central region of an optimized vortex's phase mask can be filled with a
phase profile encoding one or more additional optimized optical vortices.  In this case,
the inner limit of the first phase mask establishes the maximum outer
limit of the next, and thereby
helps to determine the projected radius of the second vortex through Eq.~(\ref{eq:radius}).
Depending on its winding number $\ell$, the secondary optical vortex can be
either larger or smaller than the first.
The optimal inner limit of the secondary vortex's annular phase mask again is determined
by Eq.~(\ref{eq:optimal}).
The number of vortices that can be projected in this way is
limited by the size and resolution of the SLM.

A typical phase mask encoding two concentric optical vortices is shown 
in Fig.~\ref{fig:setup}(a),
and the associated pattern of traps appears in Fig.~\ref{fig:setup}(b).
The two optical vortices' relative intensities can be tuned 
by adjusting their annular phase masks' inner limits.
Their relative radii can be adjusted by setting both
the outer limits and the topological charges.
The combination of relative intensity, geometry and topological charge helps to establish
the relative torques that the concentric focused rings of light exert
on trapped objects.
The resulting tunability is very helpful for controlled studies of
driven colloidal hydrodynamics.
Here we have projected two concentric vortices of topological charges
$\ell_1 = \pm 30$ and
$\ell_2 = 80$, with focused principal rings formed at 
$R_1 = 2.9~\micron$ and $R_2 = 6.4~\micron$ respectively.

In practice, the part of the input beam passing through 
featureless regions of 
the composite phase mask in Fig.~\ref{fig:setup}(a) 
propagates along the optical axis and focuses
to a conventional optical tweezer in the middle of the field of view.  
We have eliminated this
undiffracted spot with a spatial filter \cite{korda02} and displaced
the concentric
vortices by adding a phase function
\begin{equation}
  \label{eq:displace}
  \varphi_{\vec{k}}(\vecr) = \vec{k} \cdot \vecr + \frac{k_z r^2}{f},
\end{equation}
where the in-plane wavenumbers, $k_x$ and $k_y$, 
set the in-plane displacement and the axial wavenumber
$k_z = 2 \pi z / (\lambda f)$
displaces the focal plane along the optical axis \cite{liesener00,curtis02}.
Figure~\ref{fig:setup}(a) shows $\varphi(\vecr)$ without $\varphi_{\vec{k}}(\vecr)$
for clarity.
When combined with appropriate control and correction of aberrations, 
these displacements yield the comparatively circular and uniformly bright rings in
Fig.~\ref{fig:setup}(b).
The rings were imaged at low laser intensity by placing a mirror in the microscope's
focal plane and projecting the reflected light along the microscope's imaging train
to a charge-coupled device (CCD) camera (NEC TI-324AII).
Each ring can trap and circulate a single colloidal particle,
and the pair can organize multiple particles into concentric circulating rings, as shown
in the bright-field image in Fig.~\ref{fig:setup}(c).

Our samples consist of monodisperse polystyrene sulfate microspheres 
(Bangs Laboratories L030305C) 
$2a  = 0.99 ~\micron$ in diameter dispersed in a layer of water 
$18 \pm 2\micron$ thick sandwiched
between the parallel glass surfaces of a microscope slide and a \#1
cover slip.
The concentric optical vortices' axial intensity gradients 
were not strong enough to overcome radiation pressure
and trap these particles
stably in the axial direction.
Consequently, we focused the traps near the upper surface, which
prevented
the particles from escaping while permitting them to circulate 
freely around the rings.

The concentric vortices' circumferential intensity variations were
estimated by fitting the radial
intensity profile in images such as Fig.~\ref{fig:setup}(b)
to the predicted cross-section \cite{sundbeck04}.
Azimuthally-resolved intensity variations, plotted in
Fig.~\ref{fig:setup}(d),
are smooth enough that particles do not become trapped
in localized hot spots \cite{curtis03}.
We measured the individual optical vortices' transport properties by
loading a single particle into one of the rings and tracking its
motions \cite{crocker96}.
Consistent results were obtained for laser powers ranging from $P = 0.4~\unit{W}$ to
$P = 2.6~\unit{W}$, with the measured single-particle speeds scaling
nearly linearly with power.
The data from all runs were normalized by power and combined into the results
plotted in Fig.~\ref{fig:setup}(e).
Somewhat paradoxically, regions of minimum circulation speed appear
to be
correlated with the brightest regions of the
ring, where the optically-induced torque should be greatest.
This demonstrates that the intensity variations around each ring's
circumference result in attractive
optical gradient forces.
The importance of this gradient attraction 
relative to the torque exerted
by the photon's orbital angular momentum
confirms that the particles absorb a very small 
fraction of the incident photons.

\begin{figure}[t!]
  \centering
  \includegraphics[width=0.7\columnwidth]{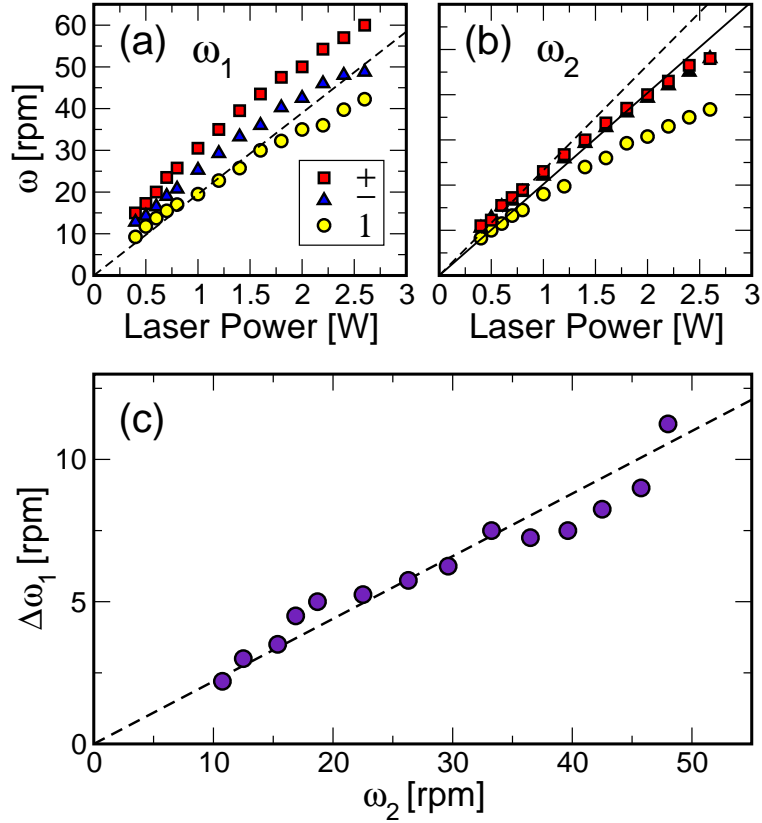}
  \caption{Colloidal particles driven by concentric optical vortices.
    (a,b) Rotation rates of the inner ($R_1$) and outer ($R_2$) 
    optical vortex:
    single particle in the system (circles), filled vortices co-rotating
    (squares) and counter-rotating (triangles).
    (c) Hydrodynamic coupling of two concentric, rotating rings of particles,
    measured by the change in the inner ring rotation rate 
    as it's direction is switched, $\Delta\omega_1$. 
    Drag force represented by the rotation rate of the outter ring $\omega_2$. 
    Dashed line is the linear fit from Eq.~(\ref{eq:drag}).
  }
  \label{fig:velocity}
\end{figure}

Despite the particles' preference for brighter regions, there is no
evidence of
circumferential
trapping, and we are justified in averaging the circulation
speed over angles to characterize the rings' overall performance.
The mean circulation speed for a single sphere on the outer and inner rings is plotted
as a function of laser power in Figs.~\ref{fig:velocity}(a) and \ref{fig:velocity}(b).
Departures from linearity in the power dependence are comparable to our measurement error
over the range, and can be ascribed to increased hydrodynamic coupling
to the upper glass surface with increasing radiation pressure.
Consistent results for both rings were obtained for $\ell_1 = +30$ and $\ell_1 = -30$.

The hydrodynamic drag on a single sphere of radius $a$
moving at speed $v$ through an unbounded fluid of viscosity $\eta$ is 
$F_\eta = -\gamma_0 \, v$, where $\gamma_0 = 6 \pi \eta a$ is the
Stokes drag coefficient.
Adding particles to one ring reduces the drag on each sphere through
the familiar drafting effect.
For $N$ spheres equally spaced around a ring of radius $R \gg a$, the
modified single-sphere drag coefficient is \cite{reichert04}
\begin{equation}
  \label{eq:draft}
  \gamma_N\left(\frac{a}{R}\right) = \gamma_0 \, \left[1 + \frac{3}{8} \, 
    \sum_{j = 2}^N \frac{a}{r_{1j}} \, \left(1 + 3 \cos \theta_{1j}\right) \right]^{-1},
\end{equation}
where $\theta_{ij} = (2\pi/N)(j - i)$ is the angular separation
between spheres $i$ and $j$, and 
$r_{ij} = R\, \sqrt{2 - 2\cos \theta_{ij}}$ is their
spatial separation.
We can use this result as a starting point to predict the filled
rings' circulation rates based on our single-particle measurements.

Coupling between the two rings should increase the drag on spheres in
counter-rotating rings, $\ell_1 = -30$, and reduce it 
for co-rotating rings, $\ell_1 = +30$.
This is indeed the case, with the effect being considerably more
pronounced for the inner ring (Fig.~\ref{fig:velocity}(a)) than
the outer (Fig.~\ref{fig:velocity}(b)).
The predictions of Eq.~(\ref{eq:draft})
are plotted as dashed lines in these figures.
Although Eq.~(\ref{eq:draft}) adequately describes the motions of the outer ring of spheres,
it substantially underestimates the hydrodynamic coupling among particles on the inner
ring.  
This might result from the nonuniform separation between spheres as
they circumnavigate the inner ring's intensity variations, 
which are more pronounced than on the outer ring, 
Fig.~\ref{fig:setup}(d) and (e).

Even this level of agreement may be coincidental, however.
The spheres in this experiment are separated from a nearby
wall by a center-to-surface distance $h$ that is not accounted for by
Eq.~(\ref{eq:draft}).
The no-slip boundary condition on the wall modifies
the spheres'  far-field flow patterns, reducing their mutual hydrodynamic
coupling \cite{dufresne00}.
At the level of the stokeslet approximation \cite{pozrikidis92}, 
the
equivalent wall-corrected result 
to $\order{(h/R)^3}$
is
\begin{multline}
  \label{eq:walldraft}
  \gamma_N^w \left( \frac{h}{R},\frac{a}{h} \right) = 
  \gw \,
  \Bigg\{ 1 + 
  \frac{3}{8} \frac{a}{R} \, \frac{\gw}{\gamma_0} \,
  \sum_{j = 2}^N \Bigg[
  \frac{(1 + \cos\theta_{1j})(1 - 3\cos\theta_{1j})}{
    \sqrt{2 - 2\cos\theta_{1j})}} \,
  \frac{h}{R} + \\
  \frac{8\cos\theta_{1j}}{(2-2\cos\theta_{1j})^{3/2}} \, 
  \frac{h^2}{R^2} +
  \frac{6(1 + \cos\theta_{1j})(5\cos\theta_{1j}-3)}{
    (2-2\cos\theta_{1j})^{3/2}} \frac{h^3}{R^3}
  \Bigg]
  \Bigg\}^{-1},
\end{multline}
where
\begin{equation}
  \gw = \gamma_0 \, \Bigg[1 - \frac{9}{16} \frac{a}{h} + 
  \frac{1}{8} \left(\frac{a}{h}\right)^3 -
  \frac{45}{256}\left(\frac{a}{h}\right)^4 -
  \frac{1}{16} \left(\frac{a}{h}\right)^5 + 
  \order{\left(\frac{a}{h}\right)^6}\Bigg]^{-1},
\end{equation}
is the single
particle's wall-corrected Stokes drag coefficient.
This still shows an overall reduction in the drag due to drafting,
but to a substantially reduced degree.
The modified functional dependence on the inter-particle separation
suggests that the spectrum of instabilities for
such symmetric configurations \cite{reichert04} also will
be modified by coupling to bounding surfaces.
A fit of Eq.~(\ref{eq:walldraft}) to the results for the outer ring, 
plotted as a solid curve in Fig.~\ref{fig:velocity}(b),
yields $h = 3a$.
This is consistent with the measured
sphere-wall separation.

Even at this low level of approximation, 
the predicted reduction in drag accounts for the
increased circulation rates when five particles are loaded onto the
inner ring and twelve onto the outer, as shown in
Fig.~\ref{fig:setup}(c).
The circulation rates on the outer ring 
are increased by 30\% (Fig.~\ref{fig:velocity}(b)),
and on the inner ring by 47\%
when the two rings are co-rotating and by 22\% when they are
counter-rotating (Fig.~\ref{fig:velocity}(a)), 
over the range of laser powers applied.

The data in Fig.~\ref{fig:velocity} also reveal that the many-body
hydrodynamic coupling in this system depends on whether the two rings of
spheres are co-rotating ($\ell_1 = +30$) or counter-rotating ($\ell_1 = -30$).
The inner ring, in particular, circulates 20\% faster in the co-rotating
configuration.
The difference $\Delta \omega_1$ in the inner ring's circulation rate
increases with laser power, but is unlikely to result from optical
interactions directly.
Rather, the trend can be ascribed to the increasing circulation rate
of the outer ring with increasing laser power, as shown in Fig.~\ref{fig:velocity}(c).
Parameterizing $\Delta \omega_1$ by $\omega_2$ is
reasonable because the outer ring's circulation rate depends only weakly on the
relative circulation direction.

We may capture this behavior semi-quantitatively by ignoring the
concentric rings' detailed structure and treating them instead
as concentric cylinders in a Couette geometry.
In this case, the drag-induced torque between the cylinders is \cite{landau59}
\begin{equation}
  \label{eq:couette}
  T = 4 \pi \eta \, H \, \frac{R_1^2R_2^2}{R_2^2-R_1^2} \,
  (\omega_2 - \omega_1),
\end{equation}
where $H$ is the effective height of the cylinders.
The change in frequency upon switching the relative circulation
direction is then
\begin{equation}
  \label{eq:drag}
  \Delta \omega_1 = (2 \omega_2) \, \left[1 + \frac{3}{2} \, \frac{a}{H} \, 
    \left(1 - \frac{R_1^2}{R_2^2}\right) \right]^{-1}.
\end{equation}
Fitting to the data in Fig.~\ref{fig:velocity} yields $H = 0.15 \, a$.
For continuous cylinders with no-slip boundary conditions, we would
expect $H = 2a$.  
The difference can be ascribed to the rings' roughly 25 percent
filling factors, which establish partially sliding boundary conditions.

This simple model's success suggests that the detailed distribution
of particles on the rings plays a minor role in establishing the inter-ring
hydrodynamic drag, although the discreteness itself is very important.
Consequently, colloidal particles in concentric optical vortices 
should provide a model experimental system
for studying how surface roughness influences hydrodynamic boundary
slip \cite{barrat99,bonaccurso03}, with the degree of surface structure being
determined by the particle size and filling factors.


We have demonstrated that optimized optical vortices can be
used to trap and circulate fluid-borne colloidal particles.
Concentric circulating rings of particles constitute a model test-bed 
for studying many-body hydrodynamic coupling in mesoscopic systems.
Improvements in the optimized optical vortices' uniformity will make
possible detailed investigations of hydrodynamic instabilities in 
driven many-particle systems.
More immediately, the system of optically driven colloidal rings
shows promise as a microoptomechanical Couette shear cell, with
potential applications in lab-on-a-chip systems.

\acknowledgments
This work was supported by the National Science Foundation through Grant Number
DMR-0451589 and Grant Number DBI-0233971.

\end{document}